# Strategic Optimization and Demand Response for Thermal Load Management in Multi-Regional Integrated Energy Systems: A Stackelberg Game Approach

Ranran Yang [1]

[1] Key Laboratory of Modern Power System Simulation and Control & Renewable Energy Technology (Northeast Electric Power University), Ministry of Education, Jilin 132012, China

**Abstract:** In the context of high fossil fuel consumption and inefficiency within China's energy systems, effective demand-side management is essential. This study examines the thermal characteristics of various building types across different functional areas, utilizing the concept of body coefficient to integrate their unique structural and energy use traits into a demand response framework supported by real-time pricing. We developed a Stackelberg game-based bi-level optimization model that captures the dynamic interplay of costs and benefits between integrated energy providers and users. This model is formulated into a Mixed Integer Linear Programming (MILP) problem using Karush-Kuhn-Tucker (KKT) conditions and linearized with the Big M method, subsequently solved using MATLAB and CPLEX. This approach enables distinctive management of heating loads in public and residential areas, optimizing energy efficiency while balancing the interests of both providers and users. Furthermore, the study explores how the proportion of different area types affects the potential for reducing heat loads, providing insights into the scalability and effectiveness of demand response strategies in integrated energy systems. This analysis not only highlights the economic benefits of such strategies but also their potential in reducing dependency on traditional energy sources, thus contributing to more sustainable energy system practices.

**Keywords:** Integrated Energy Systems, Stackelberg Game Theory, Thermal Load Management, Real-Time Energy Pricing, Demand-Side Energy Management

## 0 Introduction

In the context of increasing industrial development and escalating environmental pollution issues, China has set goals to achieve carbon neutrality and peak carbon emissions [1-3]. These goals drive the increase in the proportion of renewable energy in energy consumption to promote clean and efficient energy usage. Through multi-energy complementary strategies, integrated electric and thermal energy systems can flexibly utilize resources across different times and spatial scales to meet the demands for electricity and thermal comfort [4-5].

Several studies focus on utilizing the thermal characteristics of centralized heating networks to enhance the operational flexibility of systems, particularly focusing on the dynamic thermal performance (PDTP) of heating network pipelines and the buildings' thermal inertia (BTI) [7-10]. Considering PDTP involves accounting for the thermal losses in pipelines and the time delay of water temperature from the heat source to the heat loads [11]. Research on BTI explores the potential of buildings as thermal storage devices [12]. Literature [13] applies scale transformation to the simulation results of individual buildings, treating the average behavior of a building ensemble as the behavior of individual buildings, where all building spaces are heated simultaneously. Literature [14] establishes a planning model for energy stations in mixed-use areas based on bi-level programming, with examples focused on mixed-use areas where all nodes are mixed-load nodes, though it does not explicitly establish a characteristic model for the electrical and thermal loads of mixed-use areas. However, the aforementioned literature does not consider the differences in thermal storage capacities among buildings of different functional types nor the differences in heating demand periods [15-18].

Demand response can allocate loads as needed, enabling reasonable resource allocation. Literature [19] studies optimization strategies for load demand response under real-time electricity pricing, aiming to balance user thermal comfort with minimizing electricity costs. This strategy balances user comfort against cost reduction. Literature [20] has established a day-ahead stochastic optimization scheduling model that incorporates demand response and wind power, effectively generating optimal operational strategies for residential mixed energy systems and day-ahead scheduling plans for power systems. Literature [21] constructs a bi-level optimization model for integrated energy systems, considering the heating costs for residents and the generation costs for the system, utilizing the Karush-Kuhn-Tucker (KKT) optimality conditions to transform the nonlinear bi-level model into a Mixed Integer Linear Programming (MILP) equation, transforming the lower-level objectives into constraints for the upper-level objectives. Literature [22] proposes a strategy based on user-side demand response, combining the energy management of a single microgrid with the collaborative optimization of multiple microgrids to address the imbalance between supply and demand sides effectively. This strategy can effectively alleviate the imbalance issues while enhancing the operational cost efficiency of microgrids, with greater user participation yielding better optimization results.

Building on existing research, this paper addresses the diversity of heating demand periods for different functional types of buildings, constructing a bi-level optimization model that also considers the energy costs for users and the overall benefits for the system. By applying KKT optimality conditions, this complex problem is simplified into an MILP model. This approach allows for more effective adjustment of heating temperature thresholds across different types of buildings, thereby enhancing the system's thermal efficiency and flexibility.

## 1 Integrated Energy System

This study focuses on an integrated energy system that combines electricity and thermal energy, comprising generation units, energy conversion, and storage facilities to support the demands for electricity and thermal energy. The model described in this paper (see Fig. 1 below) depicts a multi-energy system that includes wind and solar energy as well as electricity purchased from the grid. Electricity is sourced from distributed generation and combined heat and power (CHP) facilities, while thermal energy is provided by CHP and electric boilers. The electrical and thermal networks are coupled through these facilities, allowing electric boilers to coordinate the grid integration of wind power during peak electricity and heating periods. On the demand side, through demand response mechanisms, users can adjust their load reduction or shift according to the peak and trough differences in electricity and thermal demands, thereby optimizing the overall coordination of the system and enhancing energy efficiency.

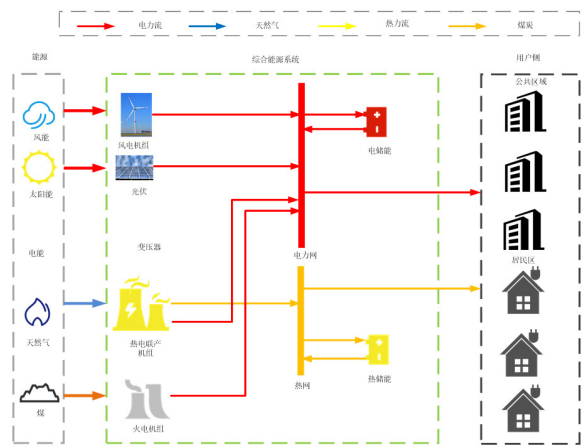

Fig.1 An integrated electric heating energy system model considering the difference of multi-district heating

## 2 Multi-regional Thermal Demand Variability Modeling

In winter, approximately 80% to 90% of thermal energy is used for heating various types of buildings. Utilizing the thermal storage characteristics of buildings can effectively mitigate the problem of wind power curtailment under the "heat-driven electricity" mode [23-25]. Buildings in different functional areas use a "body coefficient" to describe their structural differences due to their administrative properties and construction differences. Larger buildings have a lower body coefficient, while smaller buildings have a higher one [26], with public areas having a body coefficient of 0.2 and residential buildings a coefficient of 0.4.

$$Coefficient = \frac{S}{V} \tag{1}$$

Here, $S$ represents the external surface area of the building exposed to the outdoor atmosphere, and $V$ is the volume of the building.

Energy supply needs differ significantly between functional areas. For example, public buildings have cyclical and adjustable heating demands. During working hours, the indoor temperature is adjusted according to human needs to maintain comfort; at night, only a minimum temperature is maintained to prevent equipment damage due to low temperatures [27-30]. In contrast, residential buildings require uniform heating throughout the day. However, due to the subjectivity and ambiguity of thermal sensation, residential heating systems also need to be flexible. Therefore, indoor temperature thresholds are set based on time-segmented thermal comfort considerations to meet varying heat demands at different times.

## 2.1 Setting Indoor Heating Temperatures Based on Thermal Comfort in Residential Areas

The thermal load for residential heating is determined based on human comfort with temperature, which varies with different levels of activity throughout the day [31-33]. As activity increases, so does human heat dissipation, requiring stricter comfort levels regarding ambient temperature; during the night, when the body is at rest, sensitivity to temperature decreases. To meet these dynamic needs, a segmented heating strategy is adopted, allowing for moderate lowering of temperature settings to reduce thermal energy consumption and increase wind power utilization. This method not only optimizes energy use but also enhances the overall efficiency of the system [34].

Factors affecting indoor human comfort include temperature, airspeed, and relative humidity, among others. In engineering practice, a simplified formula is generally used that ignores airspeed and humidity [35-37]:

$$PMV = 2.43 - \frac{3.76(t_m - t_{in})}{M(I_{cl} + 0.1)} \tag{2}$$

In this formula, $t_m$ represents the average skin surface temperature, taken as 33.5°C; $t_{in}$ is the indoor temperature; $M$ is the metabolic rate, taken as 80 watts per square meter; $I_{cl}$ is the clothing insulation. The PMV varies within the range of [−0.5,0.5][-0.5, 0.5][−0.5,0.5], where humans do not easily notice temperature changes, and within [−1,1][-1, 1][−1,1] it also meets indoor temperature requirements for winter, hence PMV is set differently for different times: [−1,1][-1, 1][−1,1] between 8:00 and 22:00, and [−0.5,0.5][-0.5, 0.5][−0.5,0.5] during other periods.

## 2.2 Setting Indoor Heating Temperatures in Public Areas

Public areas and residential areas implement different heating strategies, mainly because the usage times in public areas are more fixed [38]. During non-operational hours, heating can be reduced to maintain a minimum "on-duty" temperature (usually set above 5°C). Thus, during operational hours in public areas, the indoor temperature is set within the [−0.5,0.5][-0.5, 0.5][−0.5,0.5] range based on the PMV index to ensure comfort; during non-operational hours, it is adjusted to stay above the on-duty temperature, effectively saving energy while meeting basic needs [39-40].

## 2.3 Heating Load Model

By maintaining indoor temperatures that satisfy resident satisfaction, the amount of heat provided to buildings can be appropriately reduced, making the demand for heat loads more flexible. According to the transient heat balance equation of buildings, the effect of the provided heat amount on indoor temperature changes is as follows [41-42]:

$$\frac{dT_t^{in}}{dt} = \frac{H_t - (T_t^{in} - T_t^{out}) \cdot K \cdot S}{c_{air} \cdot \rho_{air} \cdot V} \tag{3}$$

In this equation, $T_t^{in}$ and $T_t^{out}$ are the indoor and outdoor temperatures at time *t*, respectively; $H_t$ is the heating power; *K*, *S*, and *V* are the building's heating coefficient, surface area, and volume respectively; $c_{air}$ and $\rho_{air}$ are the specific heat capacity and density of air. From this, the relationship between heating load power and indoor temperature can be derived:

### 3 Stackelberg Game Optimization Model Considering Thermal Load Demand Variability

This study constructs a Stackelberg game optimization scheduling model where the integrated energy provider acts as the leader, and the consumers serve as the followers. This model accounts for the variable demand response of thermal loads [43-45]. The upper level of the model is controlled by the integrated energy service provider, aiming to maximize their profits by setting tiered electricity and heat prices, while the lower level allows consumers to choose their energy usage times based on these prices to minimize their energy costs and feedback their electricity usage plans to the energy provider. The structure of this model is shown in Fig. 2.

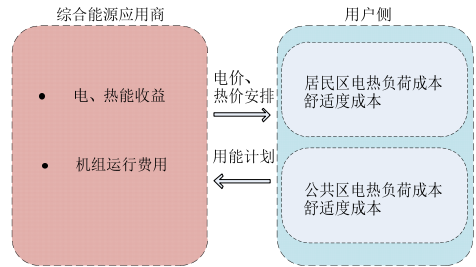

Fig. 2 Schematic diagram of game model

### 3.1 Profit Model for the Operator

The objective function of this paper is to maximize the profit after deducting the total operating costs of the system, expressed as:

$$\max C = C_{prof} - C_{opf} \tag{5}$$

Where: *C* represents the profit amount of the integrated energy provider; $C_{opf}$ represents the operational costs of the system; $C_{prof}$ represents the total sale prices of electricity and thermal energy sold by the integrated energy provider.

### 3.2 Economic Dispatch Costs of the System

The economic dispatch of the integrated energy system considers the operational costs of the units, standby costs of the units, and penalties for curtailed wind, expressed in the objective function as:

$$C_{opf} = \sum_{t=1}^{T}((\sum_{n=1}^{N} F_{CHP}^{t,n} + \sum_{m=1}^{M} F_{CON}^{t,m})\Delta T + \sum_{l=1}^{L} \pi_l (P_{l,cha} + P_{l,dis}) + \sum_{k=1}^{K} \zeta_k R_{k,t}) \tag{6}$$

Where: , represent the fuel costs of CHP and conventional thermal power units at time ttt, respectively; , represent the outputs of wind and solar power compared to actual grid power, respectively; , represent the charging and discharging power of electrical and thermal storage, respectively; is the spinning reserve capacity; is the penalty coefficient for curtailed wind and solar; is the cost coefficient for charging and discharging; represents the reserve coefficients for units and energy storage; NNN is the number of CHP units; MMM is the number of conventional thermal units; SSS includes wind and solar renewable energy sources; LLL encompasses electrical and thermal storage; KKK includes CHP units, conventional thermal units, and electrical storage.

$$F_{CHP}^{t,n} = a_{chp,n,t}(P_{chp,n,t} + c_{V,i}H_{chp,n,t})^2 + b_{chp,n}(P_{chp,n,t} + c_{V,i}H_{chp,n,t}) + c_{chp,n} \quad (7)$$

$$F_{CON}^{t,m} = a_{con,m,t}P_{con,m,t}^2 + b_{con,m,t}P_{con,m,t} + c_{con,m,t} \quad (8)$$

Where: $a_{chp,n,t}$, $b_{chp,n}$, and $c_{chp,n}$ are the fuel cost coefficients for CHP units, and $a_{con,m,t}$, $b_{con,m,t}$, $c_{con,m,t}$ are for conventional thermal units.

### 3.3 Integrated Energy Operator's Profit Amount

$$C_{prof} = \kappa_{e,t}P_{L,t} + \kappa_{h,t}(H_{L,res,t} + H_{L,pub,t}) \quad (9)$$

The above formula represents the revenue from selling electricity and heat to consumers, where $\kappa_{e,t}$ and $\kappa_{h,t}$ are the real-time electricity and heat prices at time $t$.

### 3.4 Upper-Level Model Constraints

(1) Power Balance Constraint

Electricity is used as it is generated and cannot be stored in large quantities. To meet the load requirements and ensure system stability, it is necessary to maintain the balance of electric power at every node of the system at every moment [46].

$$\sum_{n=1}^{N}P_{chp,n,t} + \sum_{m=1}^{M}P_{con,m,t} + P_{wind,t} + P_{pv,t} + P_{e,dis,t}\eta_{e,dis} - P_{e,cha,t}/\eta_{e,cha} = P_{L,t} \quad (10)$$

The thermal system must also satisfy the thermal power balance, as follows:

$$\sum_{n=1}^{N}H_{chp,n,t} + P_{eb}\eta_{eb} + P_{h,dis,t}\eta_{h,dis,t} - P_{h,cha,t}/\eta_{h,cha,t} = H_{L,res,t} + H_{L,pub,t} \quad (11)$$

Where: $P_L$ is the electrical load; $H_{L,res,t}$ and $H_{L,pub,t}$ are the heating loads for residential and public building areas, respectively.

(2) Price Constraints

To stabilize electricity and heat prices within a reasonable range and ensure the interests of the consumers are not harmed, the price range is set as follows [47-48]:

$$\sum_{t=1}^{T}\kappa_{e/h,t} = \kappa_{e/h,t,mean}T \quad (12)$$

$$\kappa_{e/h,min} \leq \kappa_{e/h,t} \leq \kappa_{e/h,max} \quad (13)$$

(3) Other Standard Constraints.

Unit output/ramping constraints and storage constraints are not detailed here but can be found in references [49-51].

### 3.5 Consumer Layer Objective Function

The consumer layer's objective function aims to minimize the cost of energy used, considering penalties for thermal comfort losses, including costs for electricity and heating.

$$\min F = \sum_{t=1}^{T}[\kappa_{e,t}P_{L,t} + \kappa_{h,t}(H_{L,res,t} + H_{L,pub,t}) + \psi \cdot (H_{cut,r,t}^2 + H_{cut,p,t}^2)] \quad (14)$$

Where: $\kappa_{e,t}$, $\kappa_{h,t}$ are the unit prices for electricity and heat, respectively, and is the coefficient for thermal comfort losses. $H_{cut,r,t}$ and $H_{cut,p,t}$ represent the thermal comfort losses in residential and public areas, respectively, derived from deviations in heat load from the most comfortable temperatures [52].

**3.6 Consumer Layer Model Constraints**

(1) Reducible Electric Load Constraint

$$P_{L,t} = P_{L0,cut,t} - \vartheta_t \Delta P_{L,cut,t} \tag{15}$$

Where: $P_{L,t}$ is the electric load after reduction at time $t$, $P_{L0,cut,t}$ is the electric load before reduction at time $t$, $\Delta P_{L,cut,t}$ is the reduced electric load, and $\vartheta_t$ is the state variable for reduced electric load [53].

(2) Reducible Thermal Load

Based on the indoor temperature constraints for heating in residential and public areas (sections 2.1 and 2.2), combined with equation (4), the boundaries for reducible thermal loads can be derived.

**4 Solution of the Game Optimization Model**

This paper considers the demand response to variability in thermal loads and uses a leader-follower game approach for solving. The solution steps are listed as follows:

**Step 1:** Input initial parameters such as wind and solar output data, unit parameters, and temperature data.

**Step 2:** Establish the upper-level optimization model for the integrated energy operator.

**Step 3:** Construct the network model.

**Step 4:** Establish the follower game model at the consumer level.

**Step 5:** Based on the Predicted Mean Vote (PMV) index, solve for the upper and lower limits of the residential heating load for each time period.

**Step 6:** For working hours, use the PMV index to establish heating load thresholds for public areas for each time period, ensuring temperatures meet the minimum 'on-duty' requirements during non-working hours.

**Step 7:** Use the KKT conditions to transform the bi-level leader-follower game model into a single-level model, and apply the Big M method to linearize the nonlinear constraints.

**Step 8:** Use MATLAB and Yalmip programming combined with CPLEX to solve the constructed model.

**Step 9:** Output the final optimization results and analyze the outcomes.

**5 Case Study Analysis**

This paper's test system employs an improved IEEE 30-bus system with two separate 6-node thermal networks, operating within the same electrical grid but as independent thermal networks, as shown in Fig. 3. The output of wind and solar power, along with outdoor temperature data, is shown in Fig. 4. The scheduling duration is 24 hours with a 1-hour time step.

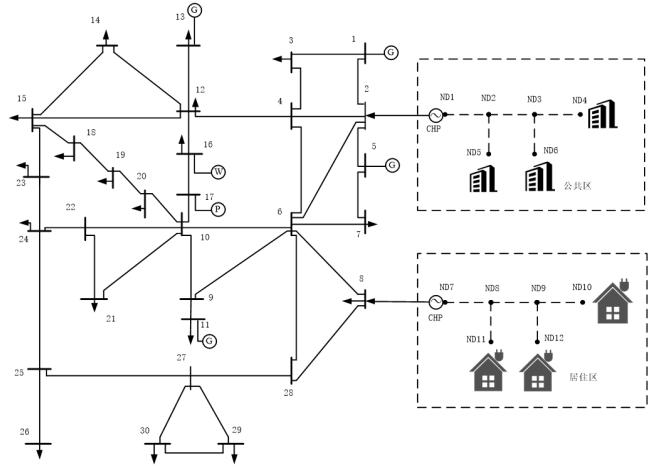

**Fig. 3 System Structure Diagram**

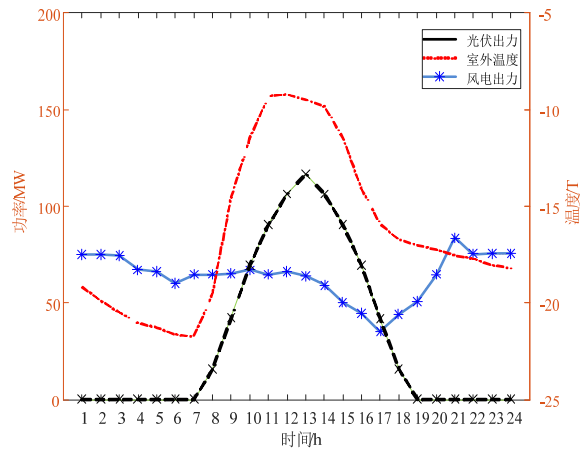

**Fig. 4 Curve of Wind and Solar Output and Outdoor Temperature**

To reasonably assess the economic impact of the proposed methods on the operation of the integrated electric-thermal energy system, the paper designs the following different modes:

- **Mode 1:** Both residential and public areas maintain indoor temperatures within the specified range of thermal comfort at all times.

- **Mode 2:** Both residential and public areas operate at optimal temperatures.

- **Mode 3:** The residential area operates according to thermal comfort temperatures, and the public area operates at optimal temperatures.

- **Mode 4:** The residential area operates at optimal temperatures, and the public area operates within the thermal comfort range.

- **Mode 5:** The residential area operates according to thermal comfort temperatures, and the public area operates differently during working and non-working hours.

**5.1 Economic Analysis:** From an economic perspective, the proposed strategies are evaluated as shown in Table 1, analyzing from both the integrated energy service provider and user dimensions in terms of their revenue and cost relationships.

**Table 1 Revenue and User Cost Analysis of Integrated Energy Applications**

| | IES Operator | | | Users | | |
|---|---|---|---|---|---|---|
| Model | Net Revenue/$ | Revenue/$ | Operating Cost/$ | Energy Cost/$ | Comfort Loss/$ | Total Cost/$ |
| 1 | 957709.2 | 1123614 | 156230.8 | 1123614 | 10196.81 | 1111134 |
| 2 | 935227.2 | 1099309 | 154635.3 | 1099309 | 0 | 1099309 |
| 3 | 904993.1 | 1065153 | 151018.7 | 1065153 | 9383.637 | 1053046 |
| 4 | 926553.9 | 1075803 | 149249.3 | 1075803 | 6270.972 | 1082074 |
| 5 | 924042.6 | 1054377 | 150474.1 | 1054377 | 20139.69 | 1074517 |

Model 5's integrated energy application provider's net revenue is at a mid-level, operating costs are relatively low, energy costs are the minimum among the proposed strategies, comfort loss is the highest, and user total costs are on the lower side. These results demonstrate the appropriateness of the non-cooperative Stackelberg strategy applied here, balancing the benefits and costs between upper and lower layer integrated energy service providers and users [54].

**5.2 Unit Output Analysis** Based on Fig. 5 below, there are clear differences in the thermal output of CHP units; among these, Model 5 has the least thermal output of CHP units, while the electrical outputs of CHP units are relatively even across all five models. This indicates that the proposed strategies significantly reduce the thermal output of CHP units, thereby reducing the use of fossil energy [55].

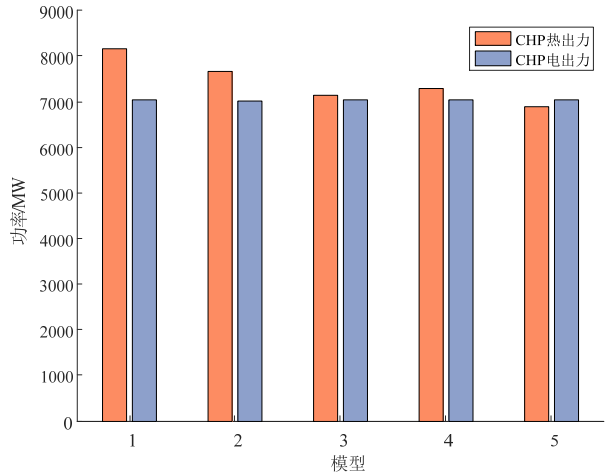

**Fig. 5 Electric and Thermal Output of Five Model CHP Units**

**5.3 Comprehensive Demand Response Analysis** (1) **Time-Shifted Electrical Loads** As controllable loads under real-time electricity pricing incentives, the time period for energy usage shifts, keeping the total electricity usage unchanged. By following the fluctuations in electricity prices, it seeks the most economical times to charge, thereby reducing user energy costs and also playing a role in peak shaving. As shown in Fig. 6, during the hours from 9:00 to 19:00, there is a reduction in peak loads. Other periods see an increase in electrical load, coinciding with high wind output, which significantly reduces curtailed wind power.

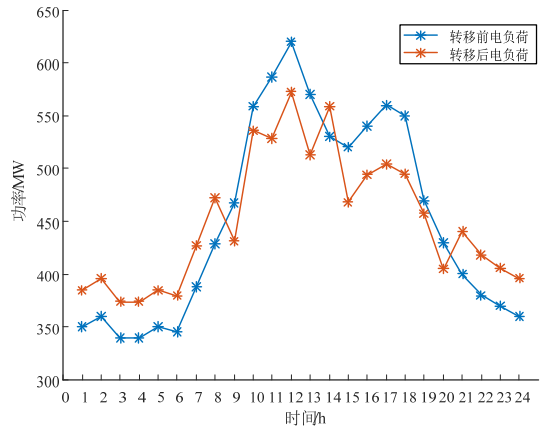

**Fig. 6 Time-Shifting Load**

(2) **Reducible Thermal Load** As shown in Fig. 7, Model 1 uses public and residential areas set according to human comfort levels and regulated by real-time heat prices, showing that when heat prices are low, reducible heat loads are small; when high, reducible heat loads are large. In Model 5, there is a negative correlation between heat prices and reducible amounts. Comparing Modes 1 and 5, the strategies proposed in this paper can more advantageously reduce heat loads, using less energy and thus reducing system energy consumption.

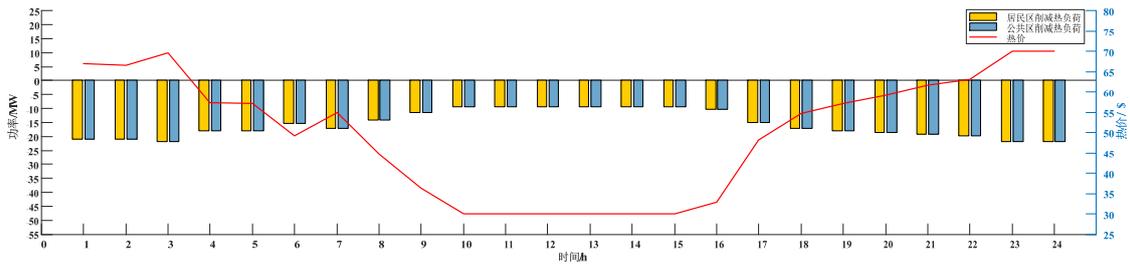

a) 模型一 居民区与公共区可削减热负荷对比及实时热价

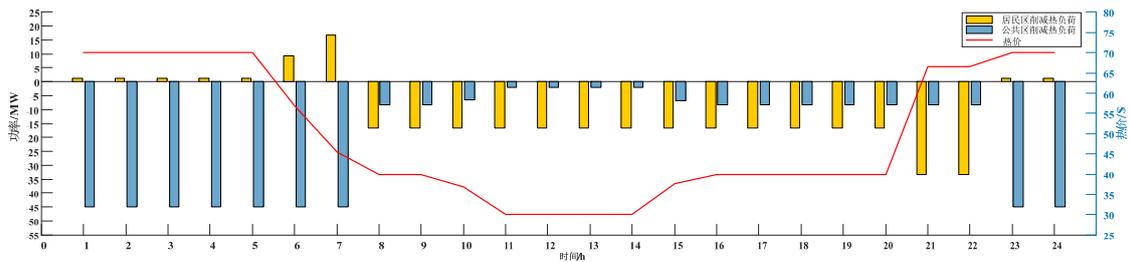

b) 模型五 居民区与公共区可削减热负荷对比及实时热价

**Fig. 7 Comparison and Real-Time Heat Price of Reducible Heat Load between Model 1 and Model 5**

As shown in Fig. 8, comparing Models 1, 3, 4, and 5, the strategies proposed in this paper achieve the greatest reduction in thermal loads. Models 3 and 4 maintain optimal temperatures in residential and public areas, respectively, but are less flexible in reducing heat loads compared to Models 1 and 5. Between 0:00 to 7:00 and 22:00 to 24:00, Models 1 and 5 achieve significant heat reductions, with Model 5 being particularly notable.

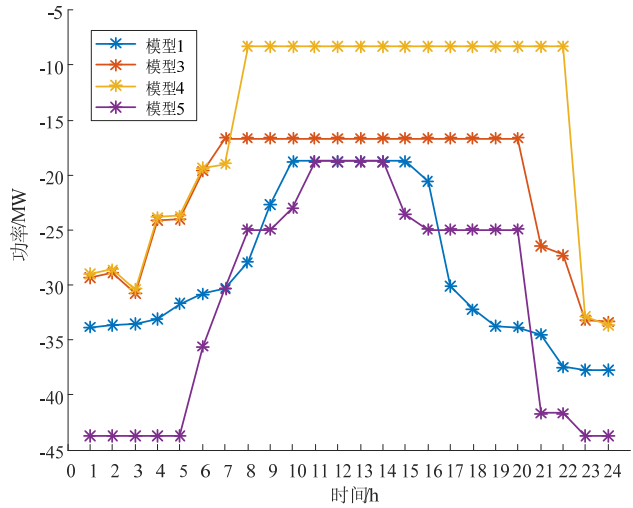

**Fig. 8 Thermal Load Reduction of Models 1, 3, 4, and 5**

**5.4 Sensitivity Analysis** As shown in Fig. 9, analyzing the proportion of heating volume in residential areas to the total heating volume at K=0.4, 0.5, and 0.6, with 0.5 where public and residential heating volumes are equal, discusses the sensitivity of the proportion. When the public area proportion is greater than the residential area, more heat load can be reduced, which suggests that increasing the public area proportion in urban planning can relatively save energy.

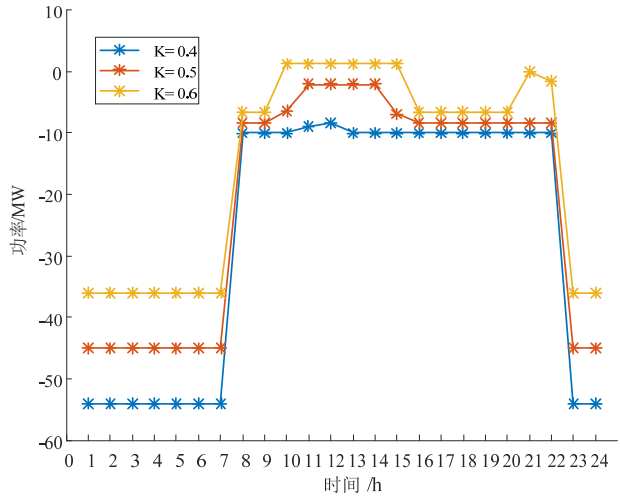

**Fig. 9 Analysis of Total Heat Reduction versus Residential Area Proportion**

**Table 2 Sensitivity Analysis of Residential Area Proportion**

| | IES Operator | | | Users | |
|---|---|---|---|---|---|
| K | Net Revenue/$ | Revenue/$ | Operating Cost /$ | Energy Cost /$ | Comfort Loss /$ | Total Cost /$ |
| 0.1 | 854678.1 | 947419.4 | 143194 | 947419.4 | 50452.68 | 997872.1 |
| 0.2 | 868351.2 | 978642.1 | 150078.1 | 978642.1 | 39787.17 | 1018429 |
| 0.3 | 899139.2 | 1011068 | 146347.5 | 1011068 | 34418.82 | 1045487 |
| 0.4 | 916758.9 | 1042311 | 152975.6 | 1042311 | 27423.95 | 1069734 |
| 0.5 | 924042.6 | 1054377 | 150474.1 | 1054377 | 20139.69 | 1074517 |
| 0.6 | 948115.8 | 1087124 | 155925.5 | 1087124 | 16917.75 | 1104041 |
| 0.7 | 995421.6 | 1127939 | 156325.7 | 1127939 | 23808.32 | 995421.6 |
| 0.8 | 1043279 | 1170367 | 155358.2 | 1170367 | 28269.84 | 1198637 |
| 0.9 | 1116348 | 1244473 | 163343.8 | 1244473 | 35218.55 | 1279691 |

From Table 2 above, it is evident that the smaller the residential area proportion, the smaller the net revenue of the integrated energy system application provider, and the smaller the energy and operating costs, but the larger the comfort loss.

## 6 Conclusion

This paper investigates the physical and energy characteristics of buildings in public and residential areas and applies these insights to the demand-side management of thermal loads. By constructing a network model based on the IEEE 30-node system and two independent 6-node thermal networks operating under a unified electrical grid, the study pioneers a comprehensive demand response Stackelberg game optimization model. This model is innovatively steered by real-time electricity and heat prices, offering new insights into energy management dynamics. The analysis yields several noteworthy conclusions:

(1) Efficiency of Stackelberg Game Model: The Stackelberg game optimization model applied facilitates a Nash equilibrium between integrated energy suppliers and users, significantly reducing the energy usage costs for users. This model demonstrates the utility of game-theoretical approaches in real-world energy systems for balancing complex multi-stakeholder interests.

(2) Advantages of Thermal Demand Response: The thermal demand response strategies, tailored to the unique physical and energy characteristics of each building type, provide distinct advantages. These strategies enable more substantial reductions in thermal loads and lessen dependency on "heat-driven electricity" approaches, which traditionally lead to the curtailment of renewable sources like wind power.

(3) Impact of Area Proportions on Energy Use: Sensitivity analysis concerning the proportion of residential areas indicates that increasing the proportion of public areas within the thermal load management system can lead to greater heat reductions, thereby improving the overall energy efficiency of the system.

Despite its contributions, this study has several acknowledged limitations: Scalability is a concern, as the models and strategies proposed are tested using a specific network configuration, and their adaptability to larger or differently configured networks remains untested. The deployment of deep reinforcement learning (DRL) could significantly enhance our model's capability to handle complex and variable operational conditions [58]. Future research prospects include expanding to larger models to validate the scalability and robustness of the findings, integrating more diverse energy sources such as geothermal or bioenergy to provide a more comprehensive view of integrated energy systems, developing advanced predictive models to better anticipate renewable power generation [59], and conducting detailed economic and environmental impact analyses of these demand response strategies at a larger scale [60-61]. By addressing these limitations and exploring these research avenues, future work can build on the foundational insights provided in this study to further enhance the management and efficiency of integrated energy systems.